# Confidence intervals with maximal average power

Christian Bartels[a]*, Johanna Mielke[b] and Ekkehard Glimm[c]

[a]*Novartis Pharma AG, Basel, Switzerland, christian.bartels@novartis.com;* [b]*Novartis Pharma AG, Basel, Switzerland;* [c]*Novartis Pharma AG, Basel, Switzerland;*

## ABSTRACT

We propose a frequentist testing procedure that maintains a defined coverage and is optimal in the sense that it gives maximal power to detect deviations from a null hypothesis when the alternative to the null hypothesis is sampled from a pre-specified distribution (the prior distribution). Selecting a prior distribution allows to tune the decision rule. This leads to an increased power, if the true data generating distribution happens to be compatible with the prior. It comes at the cost of losing power, if the data generating distribution or the observed data are incompatible with the prior. We illustrate the proposed approach for a binomial experiment, which is sufficiently simple such that the decision sets can be illustrated in figures, which should facilitate an intuitive understanding. The potential beyond the simple example will be discussed: the approach is generic in that the test is defined based on the likelihood function and the prior only. It is comparatively simple to implement and efficient to execute, since it does not rely on Minimax optimization. Conceptually it is interesting to note that for constructing the testing procedure the Bayesian posterior probability distribution is used.

## 1. INTRODUCTION

Confidence intervals are used to assess uncertainties of parameter estimates. They are related to testing null hypotheses in that in many cases the part of the parameter space outside of the confidence interval can be used as the rejection region of a test for the parameter value. Determining the confidence intervals or regions for parameters of interest is non-trivial in that (a) the confidence interval must be constructed to maintain the desired coverage (of the true, unknown parameter value) and that (b) there exist an infinite number of possible confidence intervals that maintain the desired coverage, including one sided or symmetric two sided confidence intervals. The latter implies that additional considerations are required to select or characterize confidence intervals. Even for the simple case of a binomial distribution, which is used here as example, there exist many suggestions for different exact confidence intervals starting with the Clopper-Pearson intervals (Clopper and Pearson, 1934). Clopper-Pearson intervals are based on inverting two one-sided tests, often such that each of the inverted one-sided tests guarantees a type I error of α/2. Any other choice of the type I errors for the two one-sided tests is possible as long as they sum up to α. This results in a set of possible confidence intervals that guarantee an overall type I error of α. Agresti and Min (2001) discuss and illustrate that intervals based on inverting two one-sided tests as the Clopper-Pearson intervals may not be optimal in terms of power and discuss as better alternatives intervals based on inverting one two-sided test. They summarize a few of the existing definitions.

Here, rather than proposing a testing procedure for a particular distribution and to discuss its properties beyond maintaining the desired coverage, we propose a generic frequentist testing procedure that maintains a defined coverage and is designed to be optimal in the sense that it gives maximal average power. The approach is generic in that the test is defined based on the likelihood function and the prior only. When applied to the binomial experiment, the resulting confidence intervals belong into the class of intervals obtained by inverting a single two-sided test (Agresti and Min, 2001). Using as example the comparatively simple case of a binomial experiment, the decision sets can be illustrated in figures, which should facilitate an intuitive understanding.

The approach has been introduced informally in prior publications (Bartels, 2017) and has also been applied to estimated confidence regions of a negative binomial distribution for which no established exact testing procedure exists (Bartels 2015). Here we focus on introducing the approach formally and on illustrating it for a simple well understood example. For this purpose, the notation and the flow of the arguments follow closely the work of Schafer and Stark (2009). We aim at inference on the unknown parameter of a probability distribution. We assume a parametric family of probability distributions, $\mathbb{P}_\theta$, with parameter $\theta \in \Theta$. We observed a dataset $x$ generated by a probability distribution with fixed but unknown parameter $\theta$. The dataset should be used for learning about the parameter $\theta$, or rather, we would like to identify the parameter values, which are incompatible with the data, and which we do not need to consider further in future experiments. Therefore, we aim to check the compatibility of a family of candidate parameters $\eta \epsilon \Theta$ with the data. This is a fundamental problem that exists in almost any scientific domain. The problem may be formulated as a family of statistical tests, where for each candidate parameter, $\eta$, the null hypothesis $H_{0,\eta}: \theta = \eta$ is investigated vs. the alternative hypotheses $H_{1,\eta}: \theta \in \Theta \setminus \eta$. We aim to ensure a low probability of rejecting the null hypotheses, if the parameter of the null hypothesis is equal to the true one $\eta = \theta$ (maintain type I error, coverage) and have at the same time a high probability of rejecting a null hypothesis if it is false, $\theta \in \Theta \setminus \eta$ (power). The proposed procedure gives maximal average power for the entire family of statistical tests representing all possible candidate parameters, $\eta \in \Theta$, and for all possible data generating distributions, $\theta \in \Theta$. The average is calculated using a pre-specified weighting distribution, the prior distribution. We refer to this weighting distribution as the prior distribution since in Bayesian statistics it is referred to as such, and since it is required upfront to define the test. An alternative could be to refer to it as weighting distribution for the loss function to emphasize that it enters the derivation as part of the definition of the optimality criterion used to derive the test.

The rest of the paper is structured as follows: in Section 2, we introduce the notation and the proposed methodologies. Section 3 shows the operating characteristics of the proposed approach for the example of a binary experiment comparing the performance of the test defined with an informative or non-informative prior, respectively. In Section 4, we offer some discussion points including similarities and differences relative to Shafer and Stark (2009), relations of the proposed frequentist testing procedure to Bayesian statistics and limitations of the proposed approach. Section 5 summarizes the findings.

## 2. NOTATION AND THE PROPOSED APPROACH

As in Schafer and Stark (2009), $\mathbb{P}_\theta$ denotes a parametric family of probability distributions with parameter $\theta \in \Theta$. The probability distribution is defined with respect to a probability measure $\mu$ which maps from a Borel $\sigma$-algebra that is constructed over the event set $\mathcal{X}$ to $[0,1]$. We assume that $\mathbb{P}_\theta$ with respect to this measure $\mu$ has a density which is denoted by $f_\theta$. Let $X$ be a random variable that follows the probability distribution $\mathbb{P}_\theta$ where $\theta \in \Theta$ is unknown. Realizations of $X$ are given by x.

As stated above, the proposed confidence regions are constructed aiming at maximizing the average power for rejecting null hypotheses. For that, we assume that null hypotheses $H_{0,\eta}: \theta = \eta$ are tested vs. alternative hypotheses $H_{1,\eta}: \theta \in \Theta \setminus \eta$. The decision function for this set of hypotheses is denoted by $d$ which is a measurable mapping from $\Theta \times \mathcal{X}$ into $\{0,1\}$: it has the value 0 if, based on the observations $x$, the null hypothesis is rejected and a value of 1 otherwise. The set of all decision functions $d$ is denoted by $D_\alpha$. The decision function $d$ can be used for deriving confidence regions. For that, define the candidate confidence interval set

$$C_d(x) = \{\eta \in \Theta: d(\eta, x) = 1\}, \tag{1}$$

which contains all values $\eta \in \Theta$ for which the decision function $d$ decides for the null hypothesis $\eta$ given the observed data $x$. Due to the close connection between hypothesis testing and confidence regions, this confidence region may be used as tests of the point hypotheses $H_{0,\eta}: \theta = \eta$ against all other parameter values $\Theta \setminus \eta$ with the null hypothesis $H_{0,\eta}$ being rejected, if and only if $\eta \notin C_d(x)$.

The probability that $C_d(\mathcal{X})$ covers the parameter value $\eta \in \Theta$ when in fact the random variable $X$ follows $\mathbb{P}_\theta$ is

$$\gamma_d(\theta, \eta) = P_\theta\big(\eta \in C_d(\mathcal{X})\big) = \int_\mathcal{X} d(\eta, x) f_\theta(x) \, d\mu(x). \tag{2}$$

When $\eta$ is not equal to the data generating parameter $\theta$, i.e., for $\theta \neq \eta$, $\boldsymbol{G}(\theta, \eta, d) = 1 - \gamma_d(\theta, \eta)$ is the power for correctly rejecting the null hypothesis $H_{0,\eta}: \theta = \eta$. In the other case that the data generating parameter $\theta$ is equal to the null hypothesis $\eta$, $\gamma_d(\eta, \eta)$ represents the probability of not rejecting the true data generating null hypothesis. One can also interpret it as the coverage of the corresponding confidence regions. $(1 - \alpha)$ - confidence sets are sets defined by decision functions with

$$\gamma_d(\eta, \eta) \geq 1 - \alpha. \tag{3}$$

To control type I error, we require this inequality to hold for any parameter $\eta$. Up to this point, we considered different possible data generating distributions $\mathbb{P}_\theta$ with different values $\theta \in \Theta$ separately. To define the desired power characteristics, we follow the Bayesian route and assume that the data generating parameters $\theta$ and the hypotheses $\eta$ are not fixed, but random variables. For that, define a probability measure $\upsilon$ referred to here as prior, which is defined on a Borel $\sigma$-algebra that is constructed over the event space $\Theta$. Then, the average power (over all possible parameters $\eta$) of rejecting false null hypotheses $H_{0,\eta}: \theta = \eta$ is given by

$$\boldsymbol{G}_\upsilon(\theta, d) = \int_\Theta \boldsymbol{G}(\theta, \eta, d) \, d\upsilon(\eta). \quad (4)$$

Since we do not know the true parameter $\theta$, we have to judge the power of all possible true parameters $\theta$. We do this by averaging with the prior, $\upsilon$, over the data generating parameters $\theta$

$$\boldsymbol{G}_\upsilon(d) = \int_\Theta \boldsymbol{G}_\upsilon(\theta, d) \, d\upsilon(\theta). \quad (5)$$

The optimal decision rule $d^* \in \mathcal{D}_\alpha$ that we are seeking maintains the coverage above the desired level $1 - \alpha$ (Eq. 3) for all $\theta \in \Theta$ and maximizes the average power (Eq. 5). To define the average power (Eq. 5), we followed Schafer and Stark (2009) and first averaged over different false null hypothesis (Eq. 4) and then over the data generating distributions (Eq. 5). Alternatively, as will be used below for the derivation of the optimal decision rule, the integral can be calculated by first integrating over the data generating distribution to obtain the power mixed over different data generating distributions $\theta$

$$\boldsymbol{G}'_\upsilon(\eta, d) = \int_\Theta \boldsymbol{G}(\theta, \eta, d) \, d\upsilon(\theta) = 1 - \int_\Theta \int_\mathcal{X} d(\eta, x) f_\theta(x) \, d\mu(x) \, d\upsilon(\theta). \quad (6)$$

Integrating this mixed power over different false null hypotheses, $\eta$, gives again the average power (Eq. 5).

Note that we use the same measure $\upsilon$ to average over parameters independent of whether we average over different data generating distributions $\theta$ (Eq. 5) or over null hypotheses $\eta$ (Eq. 4). This is different from Schafer and Stark (2009), who look at the minimal power $\boldsymbol{G}_\upsilon(\theta, d)$ over all possible data generating parameters $\theta$ and aim at maximizing this minimal power. This is achieved by determining the least favourable (minimax) prior $\pi$ and then using this prior to average the power over data generating distributions $\theta$. Thus, they use two different measures to average over parameter values depending on whether the parameters represent null hypotheses or data generating distributions.

The optimal decision rule $d^*$ can be derived using the the Neyman-Pearson lemma (e.g., Rüschendorf 2014 or Dudley 2003). The Neyman-Pearson lemma states that the optimal decision rule $d^*$ for deciding between two simple alternative hypotheses, e.g. between $P_\theta$ and $P_\eta$ can be expressed in terms of the ratio between the likelihoods at $\theta$ and at $\eta$. The lemma can also be applied when the alternative hypothesis is replaced by a mixture distribution $P_{mix} = \int_\Theta f_\theta(x) \, d\upsilon(\theta)$ (Rüschendorf 2014, Section 6.3) to test the null $H_0: P_\eta$ against the alternative $H_1: P_{mix}$. The power for rejecting the null hypothesis $\eta$, if the data was generated by the mixture distribution is

$$\boldsymbol{G'}_v(\eta, d) = 1 - \int_\Theta \int_X d(\eta, x) f_\theta(x) \, d\mu(x) \, dv(\theta). \tag{7}$$

The coverage for the null hypothesis $P_\eta$ is

$$\gamma_d(\eta, \eta) = P_\eta\big(\eta \in C_d(X)\big) = \int_X d(\eta, x) f_\eta(x) \, d\mu(x). \tag{8}$$

Eq (7) of the mixed power and Eq. (8) of the coverage are the same as the ones defined for our decision problem of interest (Eq. 6 and Eq. 2 with $\theta = \eta$). Thus, an optimal decision function, $d^*_v(\eta, x)$, for our problem of interest can be determined for each $\eta$ with the Neyman-Pearson lemma for the case of a mixture distribution as the alternative hypothesis. The decision function $d^*_v(\eta, x)$ determined to give maximal power $\boldsymbol{G'}_v(\eta, d)$ for each $\eta$ separately does also provide maximal average power $\boldsymbol{G}_v(d^*_v) = \int_\Theta \boldsymbol{G'}_v(\eta, d^*_v) \, dv(\eta)$. It gives maximal power since any $d^{*'}_v(\eta, x)$ with a larger average power $\boldsymbol{G}_v(d^{*'}_v)$ would require to have a higher power $\boldsymbol{G'}_v(\eta, d^{*'}_v)$ for at least some $\eta$, which contradicts the conditions of the construction of $d^*_v(\eta, x)$ with the Neyman-Pearson lemma.

The Neyman-Pearson lemma determines optimal decision rule at each $\eta$, $d^*_{v,\eta}(x) \equiv d^*_v(\eta, x)$ as $\boldsymbol{G}_v(d^*_{v,\eta}) = \sup_{d_{v,\eta} \in \mathcal{D}_\alpha} \boldsymbol{G'}_v(\eta, d_{v,\eta})$ using the likelihood ratio

$$r_v(\eta, x) = \frac{\int_\Theta f_\theta(x) \, dv(\theta)}{f_\eta(x)}. \tag{9}$$

The optimal decision rule, $d^*_v(\eta, x)$, is equal to 1 or 0 depending on whether the likelihood ratio, $r_v(\eta, x)$, is smaller or larger than a constant $c_\eta$, respectively. The constant $c_\eta$ is chosen for each null hypothesis separately as the smallest value that guarantees the desired coverage

$$\int_X d_v(\eta, x) f_\eta(x) \, d\mu(x) \geq 1 - \alpha. \tag{10}$$

The likelihood ratio $r_v(\eta, x)$ is related to the posterior $g_v(\eta, x) \, dv(\eta)$:

$$g_v(\eta, x) \, dv(\eta) \equiv \frac{dv(\eta)}{r_v(\eta, x)} = \frac{f_\eta(x) \, dv(\eta)}{\int_\Theta f_\theta(x) \, dv(\theta)} \tag{11}$$

For any given parameter $\eta$, the relation between likelihood ratio, $r_v(\eta, x)$, and posterior density, $g_v(\eta, x)$, is monotonic. Thus instead of testing whether the likelihood ratio is smaller or larger than a constant $c_\eta$, the optimal decision function may be constructed by testing whether the posterior $g_v(\eta, x)$ is larger or smaller than a constant $c'_\eta$, and choosing the constant $c'_\eta$ for each $\eta$ as the largest value that guarantees the desired coverage.

## 2.1 EXAMPLE: BINOMIAL EXPERIMENT

To help with an intuitive understanding of the effect of using an informative versus a non-informative measure $v$ (prior), the procedure is illustrated below for a binomial experiment with $n_{tot} = 100$ repetitions (see also Agresti and Min, 2001; Clopper and Pearson, 1934). The model parameter, $\theta$, is the probability of the binomial experiment, and the observation, $x$, of the result of the experiment is the number of successes and lies between 0 and 100. In Bartels (2015), the approach has been applied to the non-trivial problem of determining the two-dimensional confidence region for a negative binomial experiment.

With this example, the observations and parameters are defined on $\theta \in (0,1)$, $\eta \in (0,1)$, $x \in \{0,1,2,\ldots,n\}$ with $n = 100$. Further, we have

- the probability distribution:
$$f_\theta = Binom(x|\theta,n) = \binom{n}{x} \theta^x (1-\theta)^{n-x}$$

- the prior:
$$dv(\theta) = Beta(\theta|\alpha,\beta) = \frac{\theta^{\alpha-1}(1-\theta)^{\beta-1}}{B(\alpha,\beta)}$$

- the coverage (Eq. 2)
$$\gamma_d(\theta,\eta) = \sum_{x=0}^{n} d(\eta,x)\, Binom(x|\theta,n)$$

- the type I error
$$1 - \gamma_d(\eta,\eta) = 1 - \sum_{x=0}^{n} d(\eta,x)\, Binom(x|\eta,n)$$

- the posterior
$$g_v(\eta,x)\, dv(\eta) = Beta(\eta|\alpha+x, \beta+n-x)$$

- the mixture distribution $P_{mix}$
$$\int_\Theta f_\theta\, dv(\theta) = BetaBinom(x|\alpha,\beta,n) = \binom{n}{x} \frac{B(x+\alpha, \beta+n-x)}{B(\alpha,\beta)}$$

with $B(\alpha,\beta)$ and $\binom{n}{x}$ representing the beta function and factorial, respectively. The different average powers Eqs. (4-6) are then given by

- Average for given data generating distribution (Eq. 4):
$$\boldsymbol{G}_v(\theta,d) = 1 - \int_{\eta=0}^{1} \sum_{x=0}^{n} d(\eta,x)\, Binom(x|\theta,n)\, Beta(\eta|\alpha,\beta)\, d\eta$$

- Average for a given hypothesis (Eq. 6):
$$\boldsymbol{G'}_v(\eta,d) = 1 - \int_{\theta=0}^{1} \sum_{x=0}^{n} d(\eta,x) Binom(x|\theta,n)\, Beta(\theta|\alpha,\beta)\, d\theta$$
$$= 1 - \sum_{x=0}^{n} d(\eta,x) BetaBinom(x|\alpha,\beta,n)$$

- Overall average (Eq. 5):
$$\boldsymbol{G}_v(d) = 1 - \int_{\eta=0}^{1} \sum_{x=0}^{n} d(\eta,x) BetaBinom(x|\alpha,\beta,n)\, Beta(\eta|\alpha,\beta)\, d\eta$$

Note that for each hypothesis $H_{0,\eta}: \theta = \eta$, data points x contribute to the type I error according to the binomial distribution $P_\theta$, and they contribute to the average power $\boldsymbol{G'}_v(\eta,d)$ according to the the mixture distribution $P_{mix}$, which is a beta-binomial distribution here. The criterion to include data into the decision set (Eq. 9) according to the Neyman-Pearson lemma is the ratio of $P_{mix}$ to $P_\theta$. By using the Neyman-Pearson lemma we optimize the average power while maintaining a given type I error. Thus for each hypothesis $H_{0,\eta}: \theta = \eta$, the average power can be increased relative to type I error by including data points into the decision set, $d(\eta, x)$, with small beta-binomial probability and high binomial probability.

Two beta distributions are used to illustrate the impact of the prior measure $v$ (Figure 1): one with both the shape parameters, $\alpha, \beta$ equal to 0.5, referred to as non-informative (e.g., Kerman, 2011) in the following, and the second with both shape parameters equal to 100 and referred to as informative in the remaining text. The non-informative prior with the shape parameters of 0.5 illustrates the situation that all of the possible hypotheses are of interest (i.e., $\theta \epsilon\ [0,1]$) and that we aim for high power to reject any of them. The informative prior with the shape parameters of 100 illustrates the situation where hypotheses with parameters $\theta$ close to 0.5 are of interest. Probability parameters below 0.4 or above 0.6 have a low probability density and are essentially considered as being impossible. The limits 0 and 1 get assigned a prior probability of zero and are as such explicitly excluded from the decision problem.

**Figure 1. Informative and non-informative prior distributions**

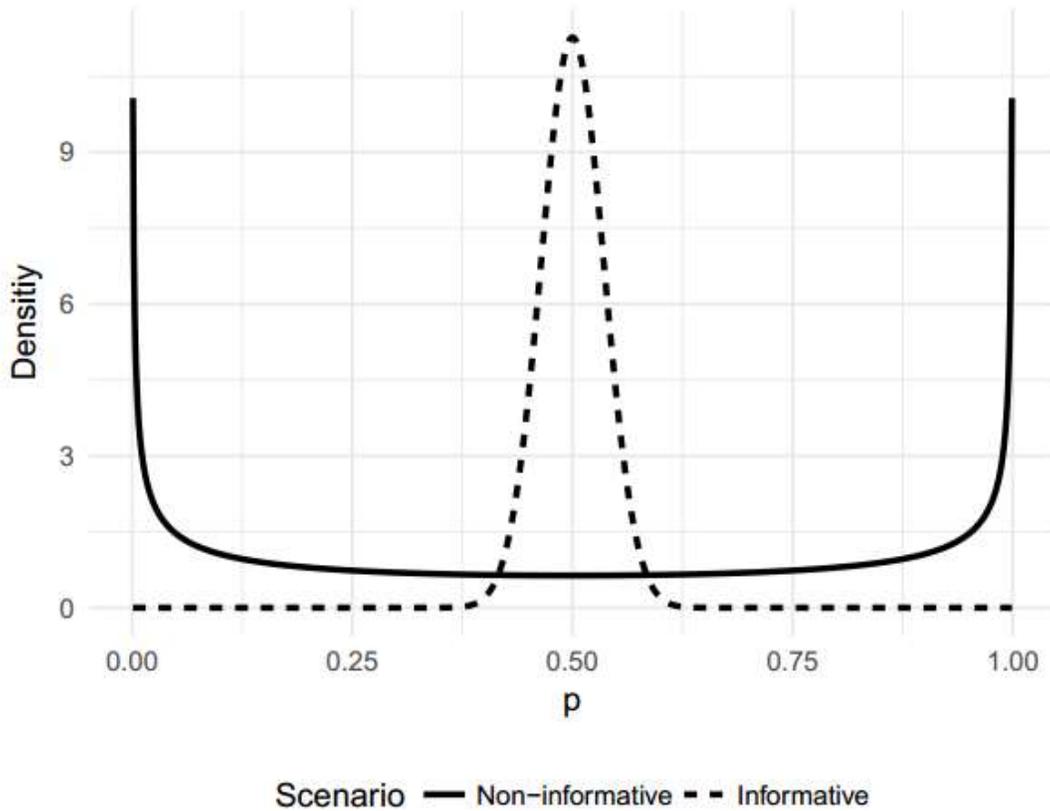

## 2.2 IMPLEMENTATION

The proposed approach is implemented in R (R Core Team 2015) using simple numerical approximations for integrals for which no closed form solution was available. The code is available as an online supplement.

The decision set (Eq. 1) is determined at 499 possible values $\eta$ equally spaced between 0.002 and 0.998. The decision function for each of the null hypothesis $\eta$ can be determined exactly. The integrals over outcomes (e.g. Eq. 10) are for the present example just sums over all possible outcomes and can be calculated as such. The posteriors (Eq. 11) are beta distributions whose densities are available in R. Integrals over parameter values (e.g. Eq. 4) used to illustrate the average power are approximated by a piecewise constant integration over the 499 parameter values used to construct the decision set.

The steps to construct the decision sets are for any null hypothesis $H_{0,\eta}$:

(1) The posterior $g_v(\eta, x)$ (Eq. 11) is determined for all possible outcomes $x$.
(2) Outcomes $x$ are included into the decision set $d(\eta, x)$ starting with those that have the largest posterior for the given parameter $\eta$.
(3) Outcomes with smaller posteriors are included until the desired coverage is reached (Eq. 10).

A generic algorithm to solve the relevant statistical integrals based on importance sampling has been proposed in Bartels (2015), but is not used here. Essentially for some actually observed data $x_0$, it is sufficient to:

(a) Sample a set of parameters $\eta_i$ that might have produced the observed data $x_0$ (same as Bayesian sampling of parameters).
(b) Sample data $x_j$ from distributions defined by the sampled set of parameters (same as Bayesian posterior predictive check).
(c) For all pairs of data and parameters, calculate the posterior density $g_v(\eta_i, x_j)$ (Eq. 11) and the contribution $f_{\eta_i}(x_i) \, d\mu(x_j)$ to the likelihood integral (Eq. 10). This can be done relatively efficiently using importance sampling. With this, go to steps 2 and 3 above to construct the decision sets.

The required calculations are similar and comparable in computational complexity to a Bayesian analysis including posterior predictive checks. Note that the observed data $x_0$ is only used to focus sampling of possible parameters and data. In the limit of an infinitely large sample of parameters and data or for simple examples as the one evaluated here, $x_0$ is not required.

## 3. APPLICATION TO BINOMIAL EXPERIMENT

Figure 2 shows the type I error of the tests for the two scenarios (uninformative vs informative prior). The type I error evaluates the situation that the null hypothesis $H_{0,\eta}: \theta = \eta$ is true. By construction, the proposed test maintains the nominal type I error rate (here: 0.05). The domain of the observations is discrete and bounded, hence the type I error can only be maintained below the desired level, unless randomized decision rules, for which $d(\theta, x)$ can assume values between 0 and 1, are used. However, in practical applications, randomized decisions are usually not accepted and therefore they are not evaluated any further here.

**Figure 2. Type I errors for different possible null hypotheses**

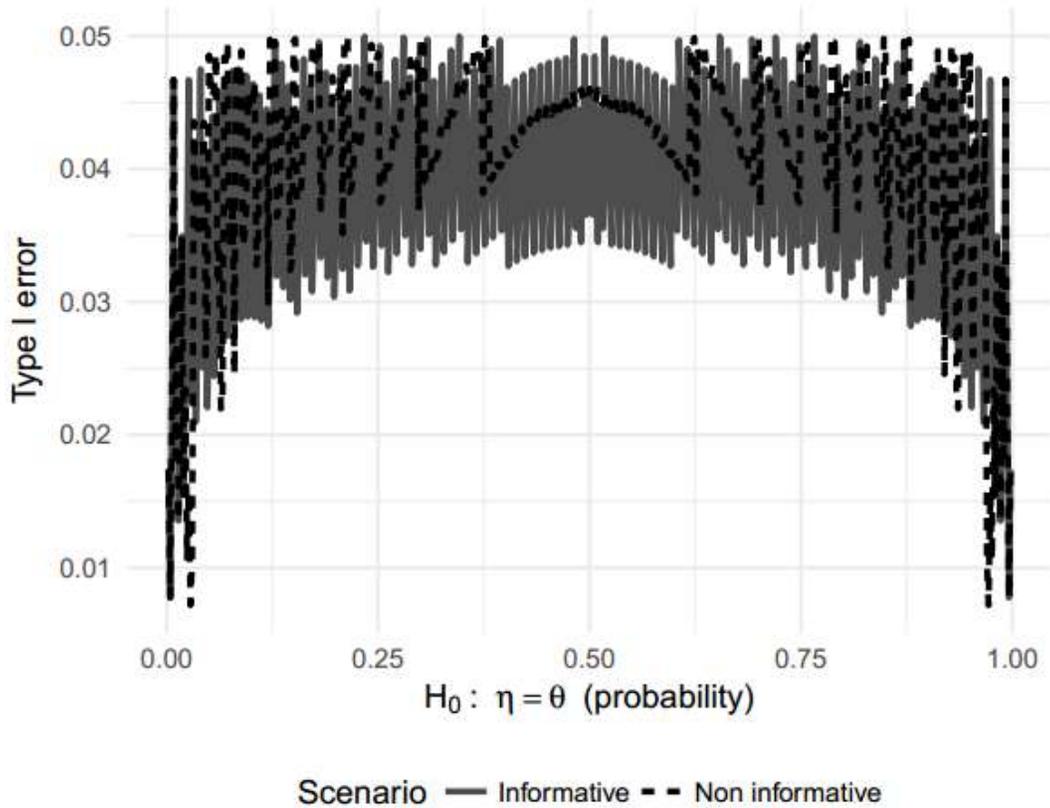

Figure 3 illustrates the power to reject null hypotheses for the two prior distributions. Three different data generating (true) distributions were chosen with $\theta = 0.5, 0.55, 0.6$, respectively. The power is illustrated dependent on the null hypotheses $H_{0,\eta}: \theta = \eta$ to be tested (x-axis). In all cases, the Type I errors are maintained. This is illustrated with the dashed lines that mark the true null hypotheses $\eta = \theta$. When comparing the three scenarios, we note that the power of rejecting other (wrong) null hypotheses depends on the parameter $\theta$ of the true sampling distribution, the parameter $\eta$ of the tested hypothesis and the reference measure (prior) $\upsilon$ used to construct the decision rule.

**Figure 3. Power of tests to reject different null hypotheses**

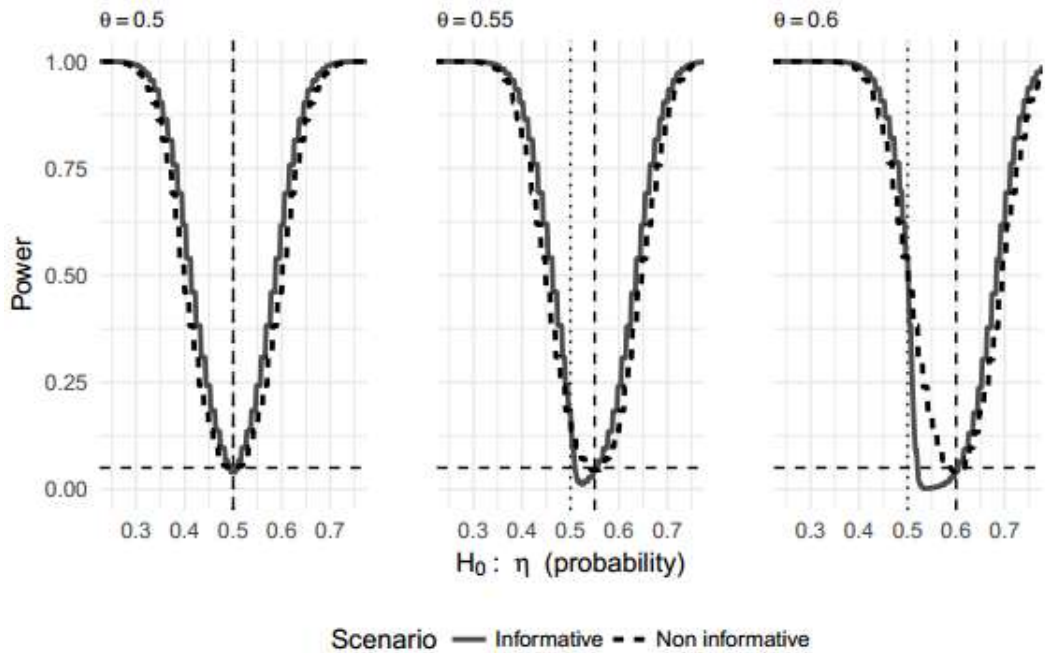

For the test constructed with the non-informative prior (dashed lines), the curves in Figure 3 are approximately symmetric and their shape does not differ much between the three selected data generating distributions. For the test constructed with informative prior, the power is increased for null hypotheses $H_{0,\eta}: \theta = \eta$ for which the data generating true distribution (parameter $\theta$, marked by dashed vertical line) and the prior distribution centered around 0.5 (marked by dotted vertical line) are in agreement. Null hypotheses corresponding to a contradiction between the true distribution and the prior (in-between the two vertical lines) have a small power and are difficult to reject. E.g., for the data generating distribution with $\theta = 0.55$, null hypotheses that are smaller than $\theta = 0.55$ but larger than the mode 0.5 of the prior (Figure 1) are difficult to reject and have a low power. Null hypotheses that are at the same time larger (or smaller) than the parameter of the data generating distribution $\theta = 0.55$ and the mode 0.5 of the prior, i.e., that have parameters larger than 0.55 or smaller than 0.5, respectively, are easier to reject and result in a power larger than the power of the non-informative test. E.g., to reject the null $H_{0,\eta=0.45}: \theta = \eta$ with the data generating distribution with $\theta = 0.55$, the power is 62% or 46% for the informative or non-informative test, respectively.

The tests were constructed to have maximal average power, when the parameter of the hypothesis $\eta$ and the data generating distribution $\theta$ follow the distribution (prior) used to construct the test. For the two tests considered here, this average power is listed in Table 1. For that, we evaluated the power of the two tests constructed with the informative and non-informative distributions, respectively, with parameters ($\theta$ and $\eta$, corresponding to the data generating distributions and null hypotheses) sampled from the informative or non-informative distribution, respectively. As designed, the informative test has a higher average power, when the informative distribution is used for the averaging than when the non-informative distribution is used, and vice-versa.

**Table 1. Average power for different sampling of hypotheses**

| Average power | Informative test | Non-informative test |
|---|---|---|
| Informative distribution of hypotheses | 0.185 | 0.154 |
| Non-informative distribution of hypotheses | 0.664 | 0.798 |

*The average power $G_v(d)$ (Eq. 5) is calculated for two different distributions, $v$, of the hypotheses and two different decision sets, $d$. The two decision sets evaluated in the two columns were constructed with the non-informative or informative prior, respectively. Similarly and shown in the rows, the non-informative or informative distributions were used to sample parameters $\theta$ of the data generating distribution and parameters $\eta$ of the null hypotheses.*

In what follows, the construction of the decision set is illustrated. For the non-informative prior, the posterior is a beta distribution with parameters $\alpha = 0.5 + x$ and $\beta = 0.5 + n - x$, which has its mode at values close to $x/n$. Observations close to the mode are those that have the largest posterior $g_v(\eta, x)$ and that are included first according to the proposed algorithm. The resulting decision sets are shown in Figure 4. The black horizontal lines illustrate the direction, in which the decision set is constructed for each $\eta$, and in which the integral (Eq. 10) is evaluated to guarantee the desired coverage. The grey vertical lines illustrate the corresponding 95% confidence interval for a given number of successes in the experiment. It consists of all hypothesized parameter values $\eta$ that cannot be rejected at a confidence level of 0.05. In Figure 4, data generating distributions (parameters $\theta$) are not explicitly illustrated; they determine the probability of observing different data. With a data generating distribution with $\theta$ close to zero, results close to zero are most likely to be observed. With a data generating distribution with $\theta$ close to one, results close to 100 are most likely to be observed. With a data generating distribution with $\theta$ equal to 0.5, results around 50 are most likely to be observed. The power, $G(\theta, \eta, d)$, discussed above is the probability of the parameter $\eta$ of the null hypothesis being outside the confidence intervals when data is sampled from the data generating distribution with parameter $\theta$.

**Figure 4. Decision set with non-informative prior**

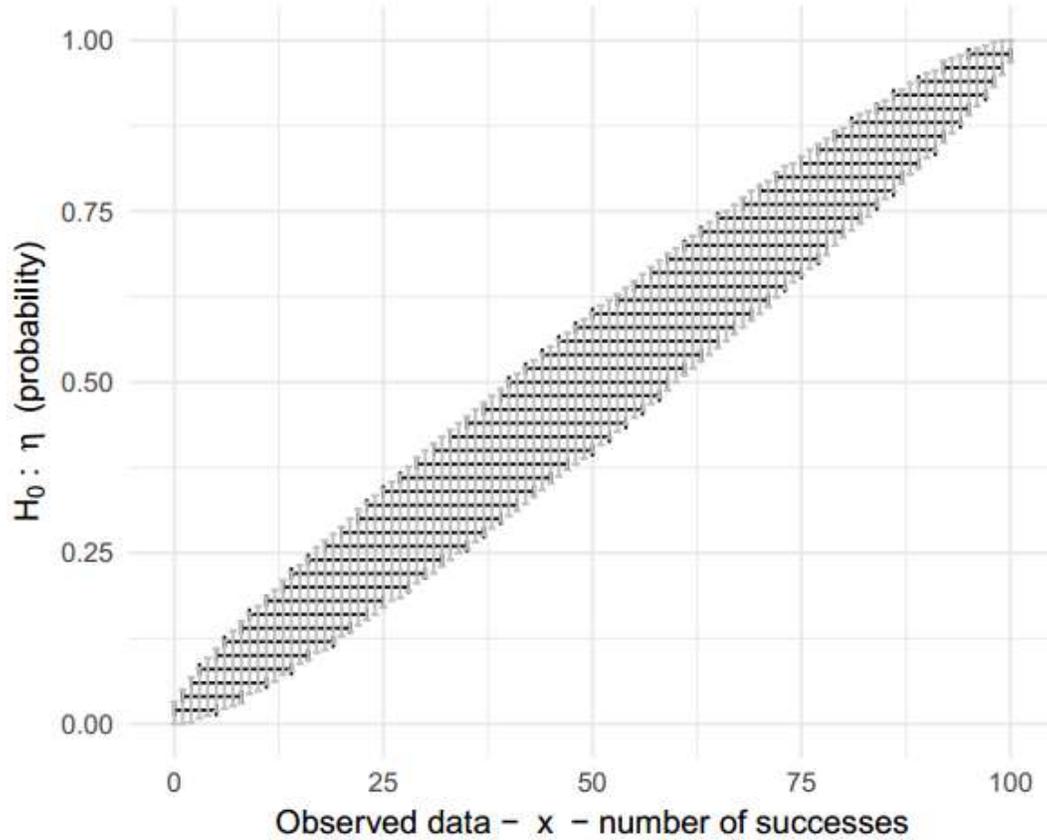

The decision sets for the case of the informative prior is illustrated in Figure 5 together with, as comparison, the decision set for the non-informative prior from Figure 4. For the informative prior, the posterior is a beta distribution with parameters $\alpha = 100 + x$ and $\beta = 100 + n - x$, which has its mode at values close to $(100 + x)/(100 + n)$. The decision set is adjusted to include, for a given observed number of successes, outcomes that are close to the mode, i.e., closer to 0.5 as compare to the set from the non-informative prior. As for the non-informative case, the coverage is ensured by including, for any true null hypothesis with parameter $\eta = \theta$, a sufficient number of outcomes into the decision set.

**Figure 5. Comparison of decision sets: non-informative versus informative prior**

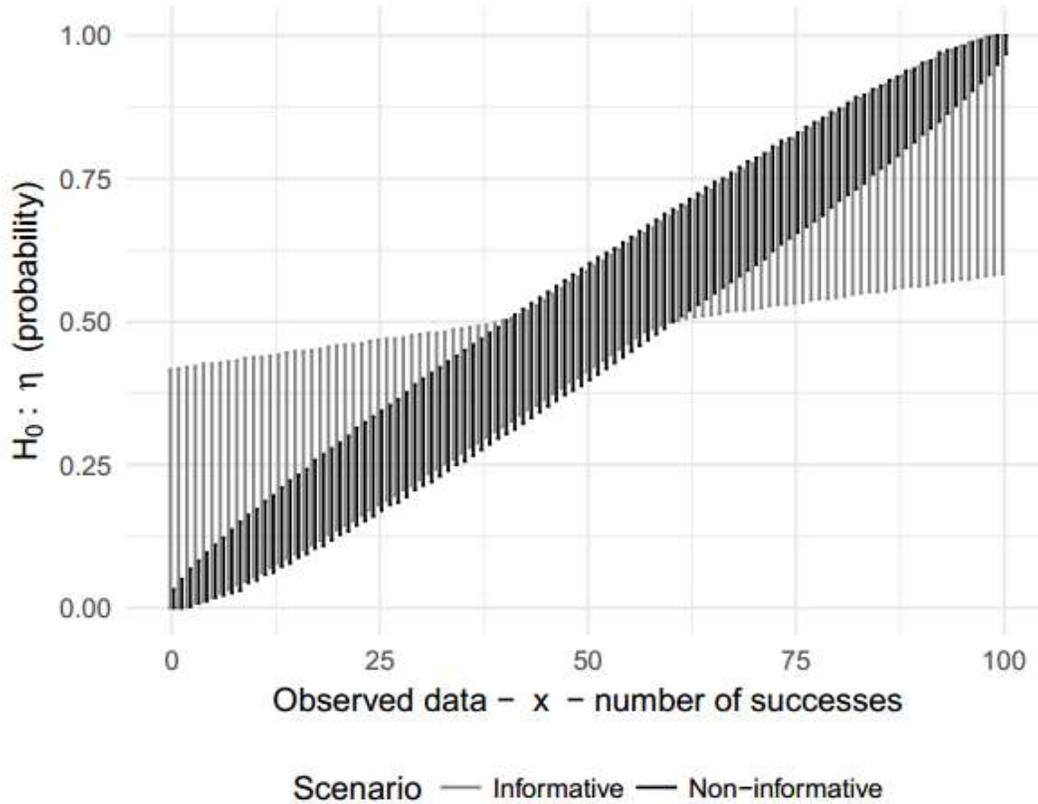

The two decision sets are different in that for observations close to 50, the informative confidence intervals are smaller than the non-informative confidence intervals. For observations closer to the extremes of 0 and 100, the informative confidence intervals are worse in rejecting hypotheses close to 0.5 but better for the hypotheses close 0 or 1. This is compatible with the power characteristics discussed in the context of Figure 3. If the observed data is consistent with the prior and lies in the middle of the prior predictive distribution, e.g., $x = 50$, the informative test is able to reject more null hypotheses.

## 4. DISCUSSION

We have proposed a frequentist procedure for testing and constructing confidence regions that is optimal in a reasonable sense, that is generic, can be implemented easily, and enables use of prior information in frequentist tests. The proposed procedure is derived from and positioned within existing, well-established mathematical statistics theory (e.g., Rüschendorf, 2014) and, as such, it is not fundamentally new. However, the approach is neither widely known, has not been published in this way as to our knowledge nor is it used in practice. This may be so, since the approach is perceived as difficult or impossible to implement, and advantages of using it are not clear. Here we have introduced the procedure as a modification of the approach proposed by Schafer and Stark (2009), and discussed and illustrated the effect of using prior knowledge in frequentist tests for a very simple example. Bartels (2015) proposed a generic implementation of the approach and applied it to the non-trivial example of determining the two-dimensional confidence region for a negative binomial experiment. The proposed approach and the example shown is related to existing work and has some limitations, e.g., it has not been established how to handle nuisance parameters in the context of the proposed approach. This will be discussed in subsequent sub-sections.

## 4.1 USE OF PRIOR INFORMATION

As to the usage of prior information, in the binary setting, we might aim to test the null hypotheses $H_{0,\eta}: \theta = \eta$ where $\eta$ gives the success probability for one experiment. Then, the approach proposed in this manuscript allows to focus on parameters close to 0.5 instead of considering all parameters with equal importance. This increases the power, if the true data generating distribution is compatible with the prior and results in confidence intervals that are more precise if the actually observed data is compatible with the prior. This comes at the cost of losing power and having larger confidence intervals, if the data generating distribution or the observed data are incompatible with the prior. E.g., as illustrated in Figure 5, when actually observing 50 successes out of 100, the test constructed with informative prior enables rejection of more parameter values than the non-informative prior. Similarly, as illustrated in Figure 3, when the data generating distribution has parameter $\theta$ equal to 0.5, the test constructed with the informative prior has higher power than the non-informative prior. This is in line with the philosophy of Mielke et al. (2018) even though their focus was primarily the Type I error rate control which is different to in the approach proposed here.

## 4.2 RELATION TO SHAFER AND STARK (2009)

Our approach as the one by Schafer and Stark (2009) aims at constructing confidence regions that guarantee a chosen coverage. This is different from the work of Habiger et al. (2013), whose proposal aims for regions that guarantee a given coverage only on average over all possible values of the parameters. The three approaches have in common that they are frequentist in nature, and that they enable to use prior information to tune the decision rule to gain more power for hypotheses that are relevant at the cost of losing power for other hypotheses that are less relevant.

There are a few, but important, differences in the proposed approach and the approach of Schafer and Stark (2009). First, instead of considering the size of the confidence regions for a parameter as the optimality criterion for the decision rule, we use the average power of rejecting null hypotheses, $H_{0,\eta}$, if they are false $\theta \in \Theta \setminus \eta$. Using the same measure to determine the size or calculating the average, these are equivalent concepts and the difference is only in nomenclature. Second, instead of aiming at the maximal average power for any possible data generating distribution (i.e. over all possible values for $\theta$) here we aim at maximal average power for data generating distributions with parameter $\theta$ sampled from a chosen distribution (the reference or prior distribution). I.e., Schafer and Stark (2009) look at the $\theta$ that has the minimal power $\boldsymbol{G}_v(\theta, d)$ and aim at a decision rule that maximizes this minimal power. Such rules are called Minimax rules. Here, we look at the average power $\boldsymbol{G}_v(d) = \int_\Theta \boldsymbol{G}_v(\theta, d) \, dv(\theta)$ and aim at a decision rule that maximizes this average power. Such rules are called Bayes rules. Both approaches make use of the Neyman-Pearson lemma to construct decision rules (Eq. 9). They differ in that here we use $dv$ as the measure for $\theta$ wheras for the Minimax approach an alternative measure $d\pi$ is constructed to identify the Minimax rule. This is done by optimizing $d\pi$ to be the least favourable prior that results in the lowest average power $\int_\Theta G_\upsilon(\theta, d) \, d\pi(\theta)$. Figure 6 shows $G_\upsilon(\theta, d)$ for the two examples presented in Section 3. A Minimax decision rule that maximizes the minimal power has similar values of $G_\upsilon(\theta, d)$ for all $\theta$. For the example with the non informative prior with both the shape parameters, $\alpha, \beta$ equal to 0.5 (Figure 1) this is approximately the case, and the optimal Minimax rule would not be too different from the one selected based on the average power. For the example with the informative prior, $G_\upsilon(\theta, d)$ has a clear minimum at $\theta = 0.5$. The least favourable prior $d\pi$ that would be constructed for the informative example would therefor further increase the weight of parameters $\theta$ close to 0.5.

**Figure 6. Average power as function of the parameters $\theta$ of the data generating distribution**

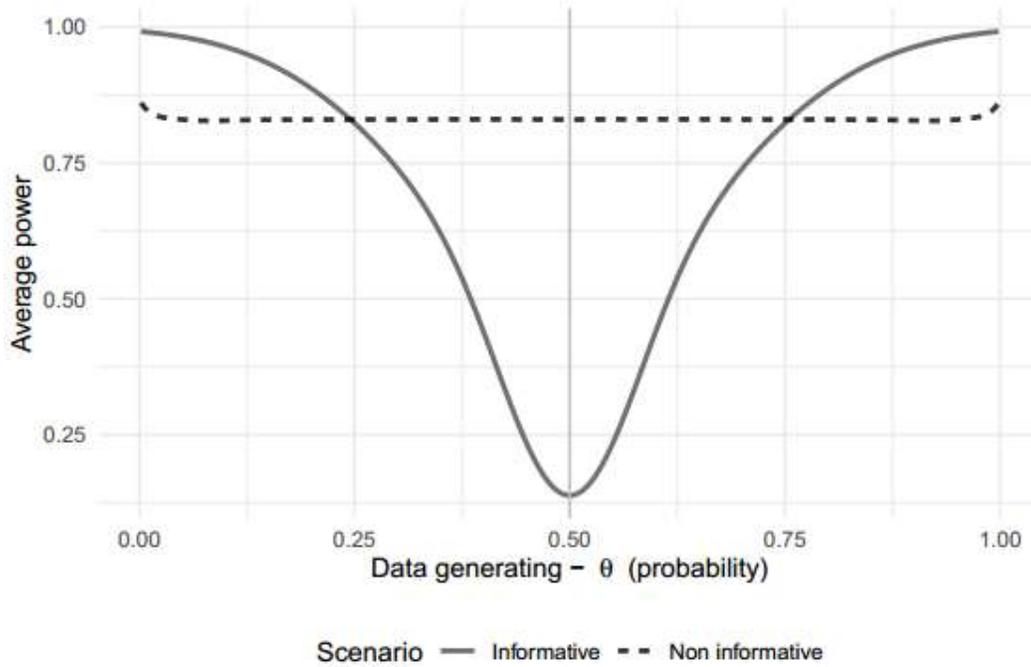

For any application, it must be decided which of the two approaches provides the desired operating characteristics under the alternative hypotheses given that both approaches guarantee a chosen coverage, $1 - \alpha$. Importantly, from an implementation point of view, the proposed approach has the advantage that it is simpler and that it does not require any Minimax optimization (e.g., Rüschendorf, 2014 or Schafer and Stark, 2009). It is also noted that any Minimax decision rule identified via its least favourable prior $d\pi$ corresponds to a rule that optimizes the average power as proposed here with the prior $d\upsilon$ used here set equal to the identified least favourable prior $d\pi$.

4.3 RELATION TO ASSURANCE

Assurance has been introduced as an alternative to classical power calculations by determining and averaging the power of a proposed experiment over a set of possible data generating alternative hypotheses instead of just choosing one (e.g., O'Hagan 2005). The idea of averaging power over data generating distributions is the same as proposed here (Eq. 6). The difference is that assurance has been introduce in the context of testing a single null hypothesis, e.g., $H_0: \theta = 0$, whereas here we consider each candidate parameter $\eta$ as defining a null hypothesis, $H_{0,\eta}$, and we test for all these hypotheses by constructing the confidence region. As a consequence the averaging of the power has to be executed also over all considered null hypotheses.

## 4.4 Relation to Bayesian inference

In the proposed approach, the dataset $x$ is assumed to be generated by a probability distribution with fixed but unknown parameter $\theta$. As such, the approach is frequentist. However, there are relations to Bayesian approaches. First, to define the criterion to judge the optimality of tests, we specify the importance of possible values of $\theta$ via a prior, which is a Bayesian idea, but it is only used to construct the frequentist test. Second, the criterion to include observations into the decision set is the posterior distribution where the prior is equal to the reference distribution used to construct the test.

In general Bayesian credible intervals and frequentist confidence intervals are different. For the particular confidence intervals proposed here, it turns out that they are similar to the credible intervals based on relative belief as proposed by Evans et al. (e.g., Evans 2016). Evans proposes to use the relative belief as a criterion to prioritize parameter values to be included into the credible intervals. The relative belief is defined as the ratio of posterior divided by the prior. With the notation used here, this is just equal to the density, $g_\upsilon(\eta, x)$, of the posterior with respect to the measure defined by the prior $d\upsilon(\eta)$. Thus, the criterion proposed here (posterior distribution or posterior density) - in a frequentist setting to include data into the decision set for any given parameter $\eta$ – is the same as the relative belief proposed in a Bayesian setting to include parameters into the credible intervals for any given observation (see Bartels 2017 for an illustration). It remains that the direction of constructing the intervals to control their size differs, even though both approaches aim at statements on parameters based on some observation $x$. In a Bayesian setting, parameters are included for a given observation to achieve the desired Bayesian coverage of the credible interval. In the proposed approach, observations are included for each parameter until the desired type I error of the test is exhausted. For this reason the two approached remain different, e.g., the Bayesian approach will, in general, fail to control type I errors, in particular for small sample sizes and discrete probability distributions.

## 4.5 Relation to other definitions of confidence intervals for a binomial experiment

Confidence intervals for a binomial experiment are well established and were discussed as part of the introduction (e.g., Clopper and Pearson, 1934; Agresti and Min, 2001). The present proposal is based on inverting a single test (Eq. 10) and is as such more related to other CI that invert a single test than to the more established Clopper-Pearson intervals.

The proposed none-informative confidence intervals (Figure 4) are similar but slightly smaller than symmetric Clopper-Pearson intervals (results not shown). This is similar as illustrated for the Blyth-Still confidence intervals in Agresti and Min (2001). The informative confidence intervals (Figure 5) are different from any possible Clopper-Pearson interval. In illustrations as used in Figures 4 and 5, the asymmetric Clopper-Pearson intervals would move all confidence intervals up or down, whereas the proposed informative test pulls the confidence intervals towards a probability of 0.5.

## 4.6 Limitations

The proposed approach is generic in that the proposed calculations require only definitions of the likelihood and the definition of the measure (prior) to calculate the average power. Also, an implementation has been proposed based on sampling similar to what is done for Bayesian analyses, which is general and should work largely independent of the chosen likelihood and prior (Bartels, 2015). Despite this, there remain limitations. Probably the most important limitation is the handling of nuisance parameters. Different approaches could be used in principle. E.g., a generic approach could be to integrate nuisance parameters out using an integrated likelihood approach (Berger et al., 1999). However, to our knowledge, it has neither been established that this would give a testing procedure that is optimal in a useful sense, nor how to implement such an approach in a generic way efficiently.

Another limitation is that the proposed approach uses the coverage and the average power as the only criterion to determine optimal confidence intervals and regions. As summarized by Agresti and Min (2001) for the case of a binomial experiment, this may not be sufficient, e.g. one may want to have one sided-tests, enforce some symmetry, or in particular, one would often want to exclude confidence intervals or regions with gaps or holes in them. Such additional criteria were not considered here, and it is not clear how they could be incorporated in a generic and efficient way other than defining the corresponding loss function and reverting to a Minimax optimization (e.g., Rüschendorf, 2014 or Schafer and Stark, 2009). Also, for the examples considered so far (binomial and negative binomial experiments), there were no gaps and holes, and maybe some mild conditions on the likelihoods and priors are sufficient to prevent them.

## 5. CONCLUSION

This paper presents an approach to construct confidence regions with optimal average power and illustrates its implementation for a binomial experiment. The resulting regions maintain type I error rates below a specified level α (equivalently: guarantee coverage above 1-α) and provide optimal power to distinguish between hypotheses $H_{0,\eta} : \theta = \eta$ where $\eta$ is sampled from a chosen distribution and $\theta$ is the parameter of the true, underlying distribution. A Bayesian posterior distribution is used to construct the test procedure and thus also the confidence region. Prior information may be used to tune the decision rule, by the choice of a distribution from which the parameters $\eta$ and $\theta$ are sampled from for the construction of the decision rule. This increases the power, if the true data generating distribution happens to be compatible with the prior. Similarly, it results in confidence intervals that are more precise if the actually observed data happens to be compatible with the prior. This comes at the cost of losing power and having larger confidence intervals if the data generating distribution or the observed data are incompatible with the prior. The proposed approach is generic, relatively simple to implement and does not rely on Minimax optimization.

## 6. ACKNOWLEDGEMENT

Johanna Mielke was supported by the Swiss State Secretariat for Education, Research and Innovation (SERI) under contract number 999754557. The opinions expressed and arguments employed herein do not necessarily reflect the official views of the Swiss Government. This project is part of the IDEAS European training network (http://www.ideas-itn.eu/) from the

European Union's Horizon 2020 research and innovation programme under the Marie Sklodowska-Curie grant agreement No 633567. We would like to thank Prof. Byron Jones for encouraging this work.